\documentclass[12pt,a4paper,dvips]{article}
\usepackage{a4p}

\usepackage{cite,mcite}
\usepackage{graphicx}
\usepackage{physics}
\usepackage{l3_title,ifthen,Lep}
\usepackage{amsmath}
\usepackage{dcolumn}

\journalname{Phys. Lett. B}
\date{October 26, 2001.}
\preprint{2001-073}
\Lep{2}
\newlength{\capindent}
\newlength{\capwidth}
\newlength{\figwidth}
\newcommand{\icaption}[2][!*!,!]{\hspace*{\capindent}%
  \begin{minipage}{\capwidth}
    \ifthenelse{\equal{#1}{!*!,!}}%
      {\caption{#2}}%
      {\caption[#1]{#2}}
  \end{minipage}}
\newcolumntype{-}{D{-}{~\mbox{--}~}{4}}
\newcolumntype{/}{D{/}{~\pm~}{2}}
\newcolumntype{+}{D{+}{~\pm~}{12}}
\newcolumntype{a}{D{a}{~\pm~}{11}}

 {\end{list}}
\def\NP{Nucl. Phys. }
\def\PL{Phys. Lett. }

\def\NIM{Nucl. Instr. Meth. }
\def\PRep{Phys. Rep. }
\def\PR{Phys. Rev. }
\def\PRL{Phys. Rev. Lett. }
\def\EURO{Eur. Phys. J. }

\def\gg{\ensuremath{\gamma \gamma}}

\def\Gggp{\ensuremath{\tilde\Gamma _{\gg}}}

\def\ra{\ensuremath{\rightarrow }}
\def\epem{\ensuremath{{\rm e}^+{\rm e}^-  }}
\def\pip{\ensuremath{\pi^+ }}
\def\pim{\ensuremath{\pi^- }}

\def\k{\ensuremath{{\rm K}}}
\def\pb{\ensuremath{{\rm pb}^{-1} }}

\def\kkpi{\ensuremath{\k \bar{\k} \pi }}
\def\kstark{\ensuremath{\k ^*\k}}

\def\etapipi{\ensuremath{\eta \,\pi^+ \pi^- }}
\def\q2{\ensuremath{Q^2 }}
\def\e{\ensuremath{\eta\,{\rm (1440)}}}

\def\fb{\ensuremath{{\rm f}_1{\rm (1420)}}}
\def\fa{\ensuremath{{\rm f}_1{\rm (1285)}}}

\def\spt{\ensuremath{P_T^2}}

\def\BR{\ensuremath{\rm BR}}

\def\a0{\ensuremath{{\rm a}_0}}
\begin{document}
\bibliographystyle{l3style}
\begin{titlepage}
\title{
\boldmath{\fa} Formation in Two-Photon Collisions at LEP
}
\author{L3 Collaboration}

\begin{abstract}
The $\eta \pi^+ \pi^-$ final state in two-photon collisions is studied with
the L3 detector at LEP, at centre-of-mass energies from 183 to 209~\GeV{}
with an integrated luminosity of 664.6~pb$^{-1}$. 
The \fa{} meson is observed and the $Q^2$ dependence of its production
is compared to different form factor models. 
The \gg{}-coupling parameter \Gggp{} is found to be $3.5\pm0.6\stat\pm0.5\sys\keV$.
The branching fraction $\Gamma\bigl(\fa\ra\a0\pi\bigr) / \Gamma\bigl(\fa\ra\eta\pi\pi\bigr)$  
is also measured.
\end{abstract}

\submitted

\end{titlepage}

\section*{Introduction}

\label{intro}
Resonance formation in two-photon interactions offers a clean environment to study the spectrum of mesonic states.
In this paper we study the reaction $\epem\ra\epem\gg\ra\epem\fa\ra\epem\,\etapipi$
in untagged two-photon collisions where
the outgoing electron and positron carry almost the full beam energy and are not detected.  
The data used for this analysis were collected with the L3 detector \cite{l3_000}
at LEP at centre-of-mass energies, $\sqrt{s}$, between 183~\GeV{} and 209~\GeV{},
corresponding to a total integrated luminosity of 664.6~\pb.

The TPC/Two-Gamma and Mark~II Collaborations observed the axial vector meson ($J^{PC}=1^{++}$) \fa{} 
in single-tag events~\cite{TPC,MRK2}.
We previously reported an indication of the formation of \fa{} in untagged events at LEP~\cite{L3}.
The \fa{} decay into $\eta\,\pi\pi$ is dominated by the two-body decay $\fa\ra\a0(980)\pi$~\cite{PDG}.
The world average for the fraction $\Gamma\bigl(\fa\ra\a0\pi\bigr) / \Gamma\bigl(\fa\ra\eta\pi\pi\bigr)$ is $0.69\pm 0.13$~\cite{PDG},
although some experiments observed only the $\a0\pi$ channel~\cite{WA76,WA102}.

In the present analysis, the formation of \fa{} is studied as a function of the transverse 
momentum squared of the \etapipi{} system, \spt{}.
To a good approximation, $\spt=\q2$ where \q2{} is the maximum virtuality of the two photons.
Production of a spin-one resonance is suppressed for real photons, according to the 
Landau-Yang theorem~\cite{Landau}.
An axial vector state can be produced in collisions of transverse-scalar virtual photons
as well as of transverse-transverse photons, when one of them is highly virtual~\cite{Schuler}. 
The \gg{}-coupling parameter \Gggp{} is defined as~\cite{TPC}:
\begin{displaymath}
\Gggp=\lim_{\q2 \to 0} \frac{M^2}{\q2}\Gamma_{\gg^*}^{\rm TS}\>,
\end{displaymath}
where $M$ is mass of the resonance and $\Gamma_{\gg^*}^{\rm TS}$ is the partial width
for the transverse-scalar photon-photon interaction.

The cross section for the formation of an axial vector meson in two-photon collisions
is described~\cite{Schuler} by:

\begin{equation}
\sigma_{\gg\ra{\rm R}} = 24\pi\frac{\Gggp\Gamma}{(W^2-M^2)^2+\Gamma^2M^2}\Bigl(1+\frac{\q2}{M^2}\Bigr) \tilde F^2(\q2)  \>,
\label{eq:ggsig}
\end{equation}

\noindent
where $W$ is the two-photon effective mass and $\tilde F$ is an effective form factor. 
The \q2{} dependence of the resonance formation can be derived~\cite{Schuler}  
using a hard scattering approach~\cite{Brodsky} and the
form factor written as:

\begin{equation}
\tilde F^2(\q2) = \frac{\q2}{M^2} \Bigl(1+\frac{\q2}{2M^2}\Bigr)\frac{2}{(1+\q2/\Lambda^2)^4} \>,
\label{eq:ffschuler}
\end{equation}

\noindent where $\Lambda$ is a parameter whose value is expected to be close to the resonance mass~\cite{Schuler}.

Previous analyses~\cite{MRK2,TPC} used the form~\cite{Cahn} 

\begin{equation}
\tilde F^2(\q2) = \frac{\q2}{M^2} \Bigl(1+\frac{\q2}{2M^2}\Bigr)\frac{2}{(1+\q2/M_\rho^2)^2} \>,
\label{eq:ffcahn}
\end{equation}

\noindent
where $M_\rho$ is the mass of the $\rho$-meson. 
The last factor is the $\rho$~pole in the vector dominance model (VDM). 
The second factor of Equations (\ref{eq:ffschuler}) and (\ref{eq:ffcahn}) includes the contributions 
from transverse-scalar and transverse-transverse photons respectively.
Both models are compared to our data.

\section*{Monte Carlo Generators}

Two Monte Carlo generators are used to describe two-photon resonance formation:
EGPC~\cite{Linde} and GaGaRes~\cite{Gulik}.

The EGPC Monte Carlo describes the two-photon process as the product of the 
luminosity function for transverse photons \cite{Budnev} and the resonance production cross section. 
The decay of the resonance is generated according to Lorentz invariant phase-space.
A Monte Carlo sample of the \fa{} meson is generated with $M=1.282\GeV$ and full width $\Gamma=0.024\GeV$~\cite{PDG}, 
for $\sqrt{s} = 189\GeV$. 
The events are passed through the L3 detector simulation
based on the GEANT~\cite{GEANT} and GEISHA~\cite{GEISHA} programs.
Time dependent detector inefficiencies, as monitored during the data taking period, are also simulated.
This sample is used to obtain the selection efficiency. 

The GaGaRes generator uses the exact matrix element for resonance production, 
$\ee \ra \ee \fa$~\cite{Schuler}. 
It describes the \q2{} dependence of axial vector meson production, 
according to the form factor~(\ref{eq:ffschuler}), 
and is used for comparison with the experimental cross section.
The \q2{} distribution does not depend on $\sqrt{s}$ for the energy range investigated.

\section*{Event Selection}

\begin{figure}[p]
\begin{center}
\includegraphics[width=\figwidth ]{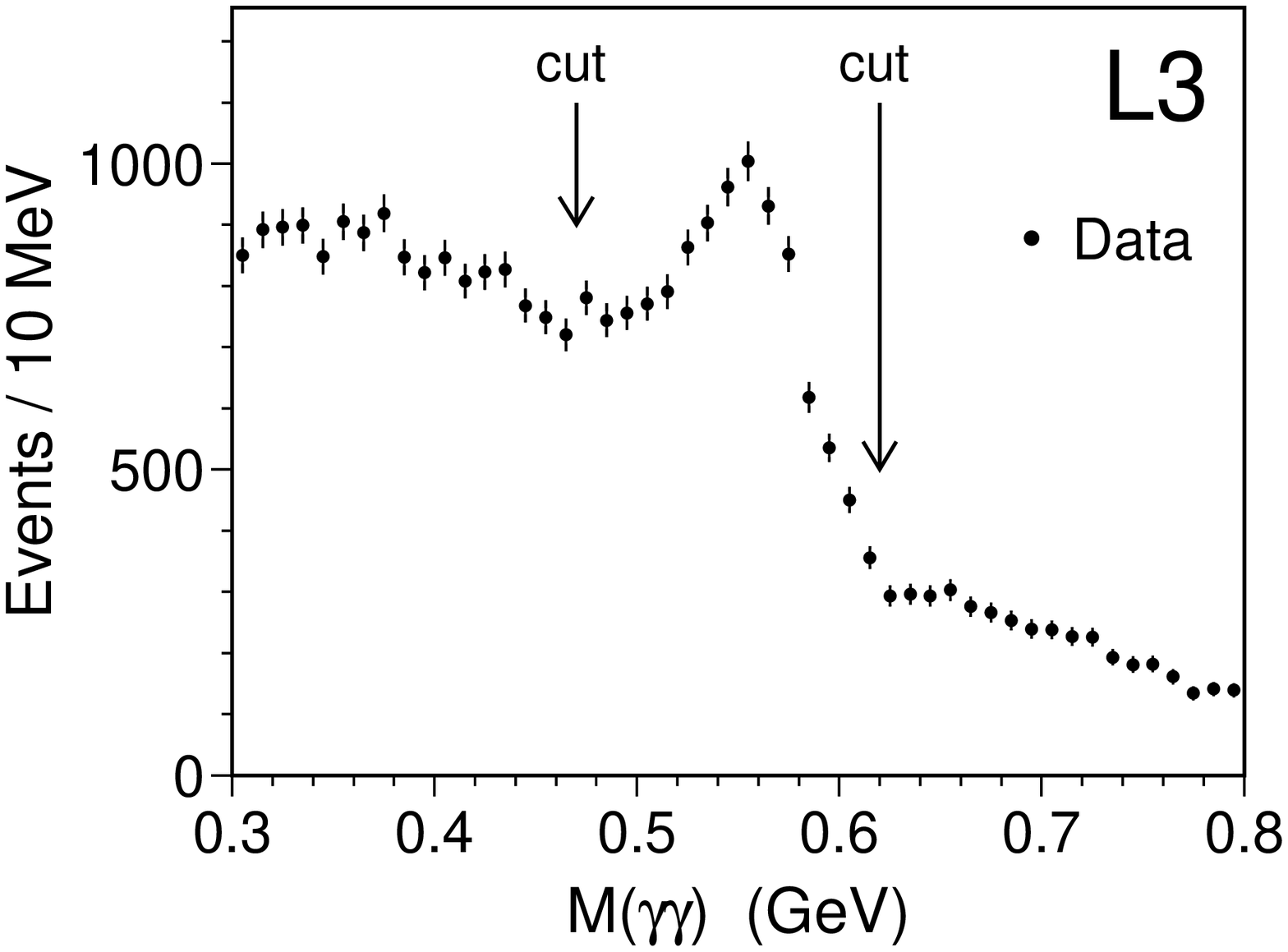}
\end{center}
\caption{The $\gg$ effective mass spectrum for events
with a   $\gg\pip\pim$ effective mass less than 2~\GeV{}.
}
\label{fig:etamass}
\end{figure}

Events from the process $\epem\ra\epem\,\etapipi$, where only the decay $\eta \ra \gamma \gamma$  is considered,
are selected by requiring two particles of opposite charge and two photons, 
since the scattered electrons go undetected at very small polar angles.
A charged particle is defined as a track in the central detector with at least 12 hits, 
coming from the interaction vertex within three standard deviations 
both in the transverse plane and along the beam axis.
The pions are identified by the $dE/dx$ measurement, requiring a confidence level greater than~1\%.
A photon is defined as a cluster in the electromagnetic calorimeter of energy greater than 0.1~\GeV{}
and with no track around 0.2~rad from its direction.
Photons in the polar angular range $0.21 < \theta < 2.93\,{\rm rad}$ are considered.
The most energetic of the two photons must have energy greater than  0.24~\GeV{}. 
A clear $\eta\ra\gg$ signal is seen in the two-photon effective mass spectrum, Figure~\ref{fig:etamass},
where $\eta$ candidates are defined by the cut (0.47$-$0.62)~\GeV{}.
The asymmetry of the two limits relative to the $\eta$ mass, $0.547\GeV$~\cite{PDG}, is due to 
a low energy tail of photon energy deposition in the electromagnetic calorimeter.
To improve the \etapipi{} mass resolution, a kinematic fit, constrained to the $\eta$ mass, is then applied.

\begin{figure}[p]
\begin{center}
\includegraphics[width=\figwidth ]{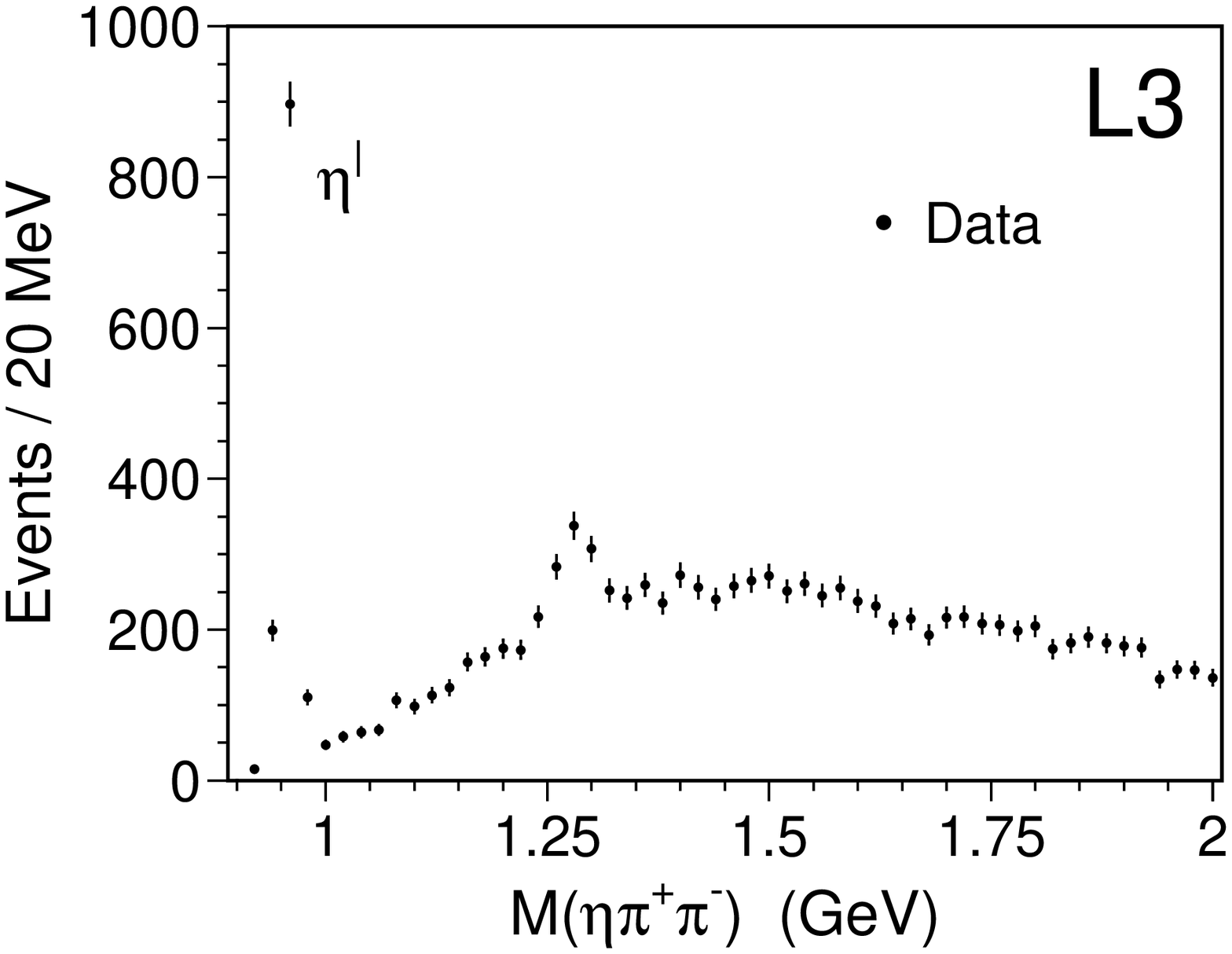}
\end{center}
\caption{The $\etapipi$ effective mass spectrum for the selected $\epem\ra \epem\etapipi$ events.
}
\label{fig:e2pi}
\end{figure}

The selection results in 11254 events with a \etapipi{} mass below 2~\GeV{}.
The $\etapipi$ mass spectrum is shown in Figure~\ref{fig:e2pi} and presents a clear peak of 
the $\eta '(958)$ resonance near threshold and a peak between 1.25~\GeV{} and 1.35~\GeV{},
which we associate with the \fa{} meson.

\section*{Results}

\subsection*{\boldmath{\fa} Formation}

\label{sect:spectra}
\begin{figure}[p]
\begin{center}
\includegraphics[width=\figwidth ]{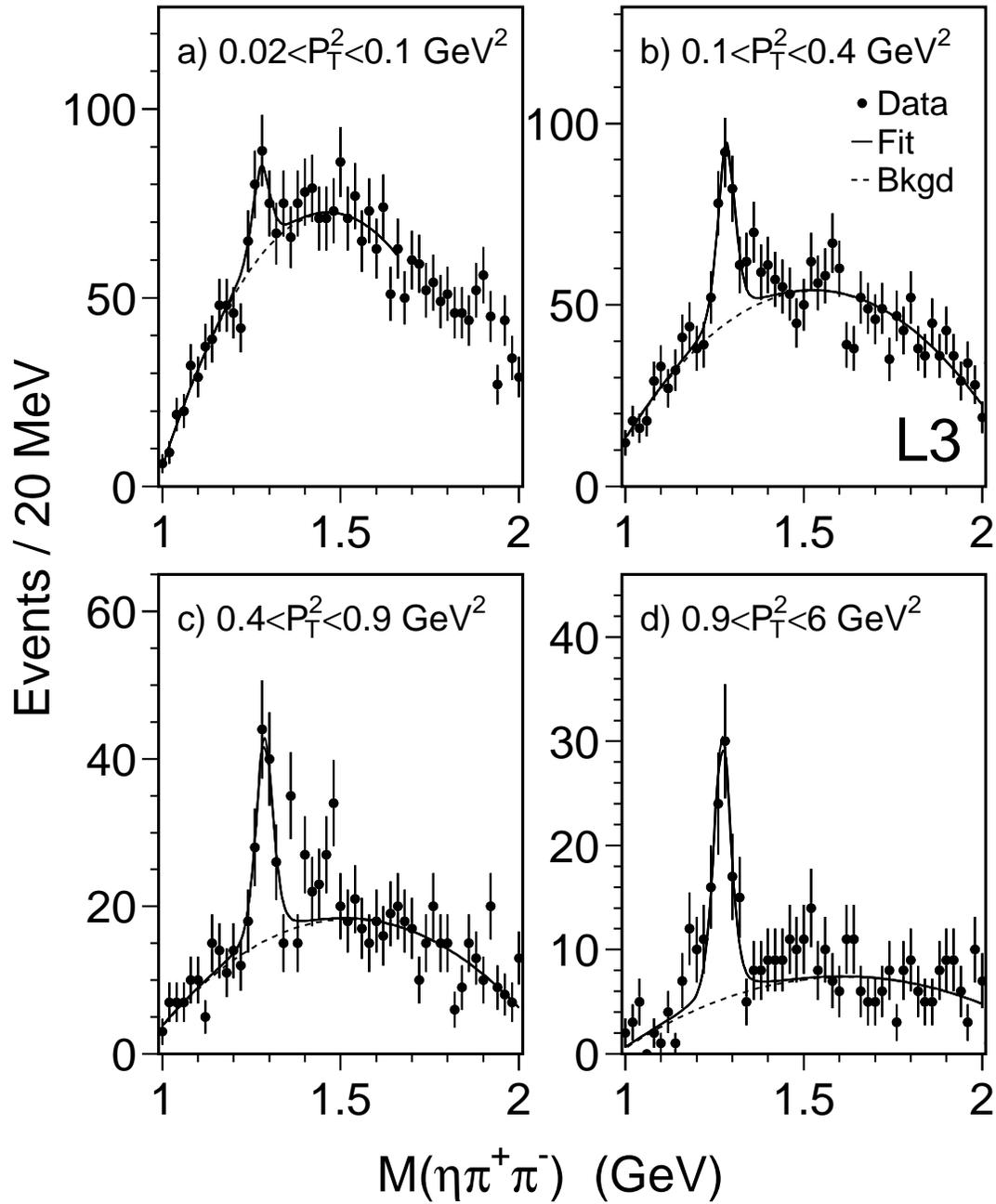}
\end{center}
\caption{The effective $\etapipi$ mass spectra for different $\spt$ bins. 
Fits of a resonance on a second order polynomial background are superimposed to the data.
}
\label{fig:f1bins}
\end{figure}

To study the formation of the $\fa$ meson, 
the data are subdivided into four \spt{} intervals, as shown in Figure~\ref{fig:f1bins} and listed in Table~\ref{tab:ptbins}.
Each spectrum is fitted with a resonance plus a background function.
The resonance is described by the convolution of a Breit-Wigner of width $\Gamma=0.024\GeV$~\cite{PDG},
with a Gaussian resolution function of width 0.018~\GeV{}, estimated with Monte Carlo.
The background is a second order polynomial.
The fit results are listed in Table~\ref{tab:ptbins}, 
the mass values obtained in the four intervals are compatible within
statistics with the mass of \fa{}, $1.2819\pm 0.0006\GeV$~\cite{PDG}.

Besides the \fa{} peak, Figures~\ref{fig:f1bins}b$-$d present a structure at masses between 1.4~\GeV{} and 1.5~\GeV{}.
This structure has variable mass and shape in these \spt{} intervals and almost disappears in the 
total spectrum for $\spt>0.1\GeV^2$, shown in Figure~\ref{fig:a0f1}a. 
Previously, the \fb{} was observed in this mass region, but only in the \kkpi{} final state, 
decaying dominantly into \kstark{}~\cite{PDG}.
A similar fluctuation in the \etapipi{} final state in the $(1^{++})$ wave was also reported~\cite{WA102}
and interpreted as an interference effect with \fb{}, decaying to $\a0\pi$.
This structure is not considered further in this letter.

The partial cross sections $\Delta\sigma$ for each \spt{} range are calculated according to: 
$$ \Delta\sigma = \frac{N}{\epsilon\, {\cal L}_{ee} \BR}\>, $$ 
where $N$ is the number of events corresponding to the peak, 
the overall efficiency, $\epsilon$, is the product of the selection
efficiency,  
obtained from Monte Carlo, and the trigger efficiency, evaluated using data.
${\cal L}_{ee}$ is the total integrated luminosity.
The trigger efficiency varies from 46\% to 40\% in the \spt{} range
from 0.02 to $6.0\GeV^2$. 
The branching ratio $\BR=0.1396$ includes $\BR\bigl(\fa\ra\eta\pi\pi\bigr)=0.528\pm 0.045$~\cite{WA102},
$\BR(\eta\ra\gg)=0.3933$~\cite{PDG} and the isospin factor
$(\pip\pim)/(\pi\pi)=2/3$. 

Table~\ref{tab:ptbins} lists the values of $\epsilon$ and $\Delta\sigma$.
The overall efficiency is found to be independent of $\sqrt{s}$.
Systematic uncertainties on $\Delta\sigma$ are presented in Table~\ref{tab:syst}.
They include the uncertainty due to Monte Carlo statistics and trigger behaviour,
the uncertainty due to background subtraction, estimated with variation of the fit ranges
and the uncertainty from event selection. 
The last is estimated by varying the $\eta$ mass range and 
the energy threshold for the most energetic photon.

\subsection*{\boldmath{\q2} Dependence}

\label{sect:q2}
\begin{figure}[p]
\begin{center}
\includegraphics[width=\figwidth ]{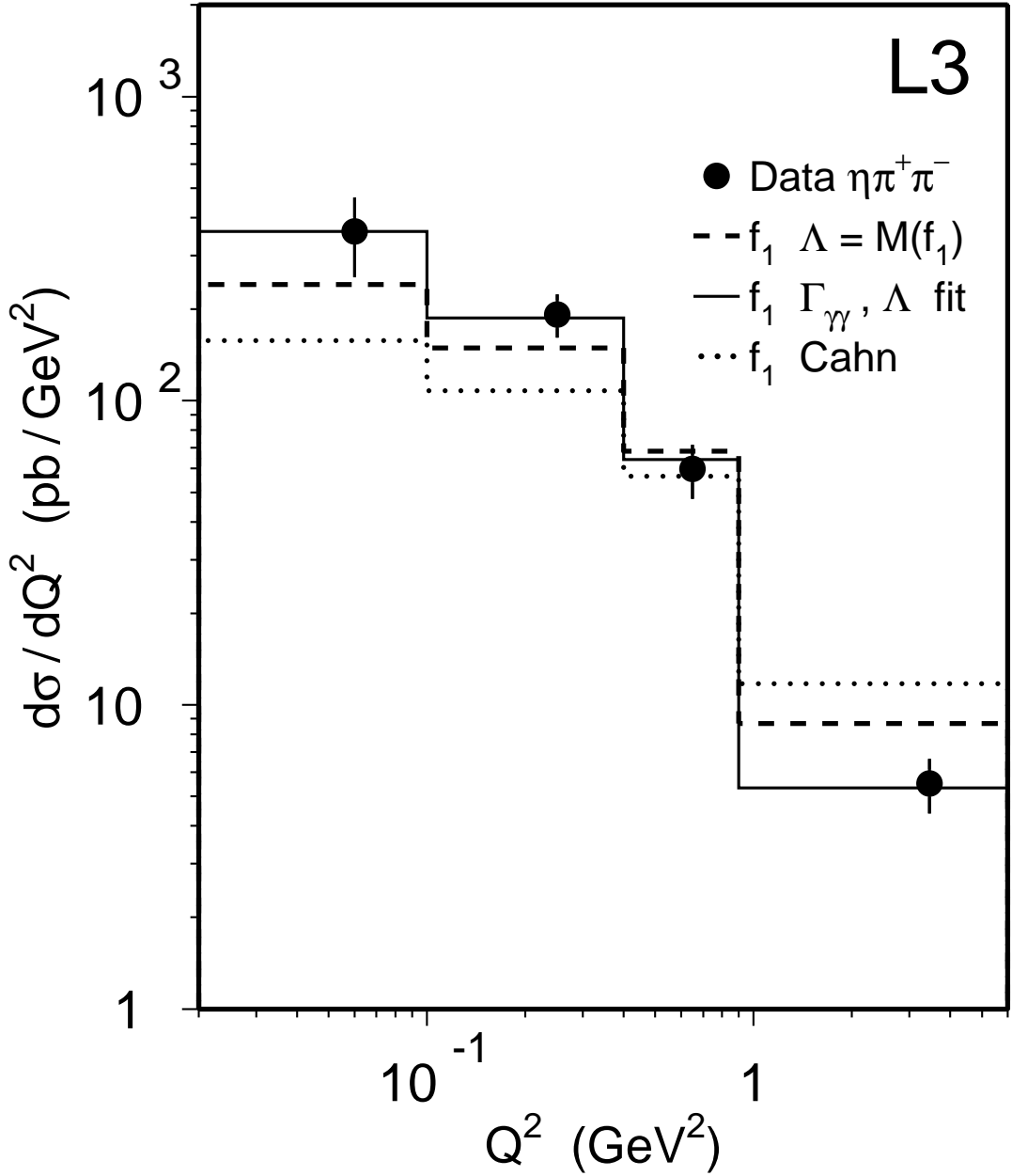}
\end{center}
\caption[]{Experimental differential cross section $d\sigma / d\q2$ compared to calculations of the 
GaGaRes Monte Carlo (dashed line) and to the calculations of Cahn~\cite{Cahn} (dotted line).
The full line is a fit of the data with the GaGaRes model, with $\Lambda$ and $\Gggp$ as free
parameters.
}
\label{fig:gagarf1}
\end{figure}

The experimental differential cross section of \fa{} production
as a function of \q2{} is presented in Figure~\ref{fig:gagarf1} and 
compared to the GaGaRes Monte Carlo prediction.
First, the mass parameter $\Lambda$ in the form factor of Equation~(\ref{eq:ffschuler})
is fixed to the resonance mass, $M=1.282\GeV$.
Normalising the Monte Carlo histogram to the experimental cross section 
in the measured interval $0.02 \le \spt \le 6.0\GeV^2$, a confidence level of 2\% is found. 
A fit of the GaGaRes prediction is then performed, where $\Lambda$ and \Gggp{} are free parameters. 
It gives:
$$\Lambda=1.04\pm 0.06\pm 0.05\GeV\>,$$
$$\Gggp=3.5\pm 0.6\pm 0.5\keV\>, $$
with a confidence level of 91\% and correlation coefficient $-$0.89.
The first uncertainty quoted is statistical and the second is systematic. 
The uncertainty on \Gggp{} includes the uncertainty on $\BR\bigr(\fa\ra\eta\pi\pi\bigl)$.

By using the fitted values of $\Lambda$ and \Gggp{}, 
we extrapolate the measured cross section to the full \spt{} range with GaGaRes, 
obtaining the value:
$$\sigma\bigl(\epem\ra \epem\fa\bigr)=155\pm 14\pm 16\>{\rm pb}\>,$$
where the first uncertainty is statistical and the second is systematic.
This value refers to a luminosity averaged $\sqrt{s}$ of $196.6\GeV$.

We also compare the experimental results to the predictions, obtained with the formalism of Reference~\citen{Cahn}, 
using the form factor defined in Equation~(\ref{eq:ffcahn}).
Normalising the prediction to the experimental cross section, a confidence level below $10^{-9}$ is found.
The incompatibility of the differential cross section shapes is evident in Figure~\ref{fig:gagarf1}.

\subsection*{\boldmath{\fa\ra\a0(980)\pi} Branching Fraction}

\begin{figure}[p]
\begin{center}
\includegraphics[width=\figwidth ]{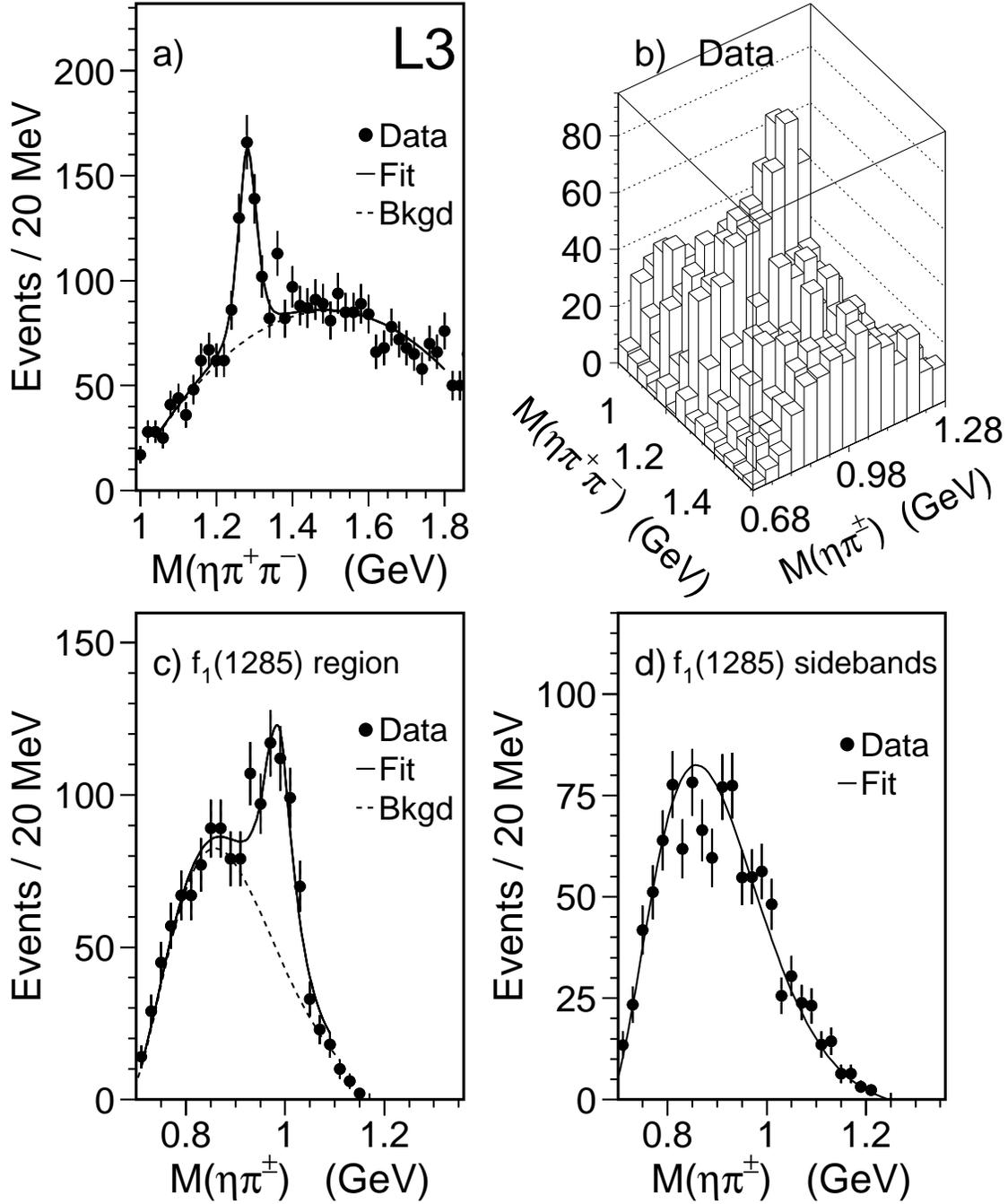}
\end{center}
\caption[]{Search for the $\fa\ra\a0(980)\pi$ decay mode.
a)~$\etapipi$ mass spectrum,
b)~masses of both $\eta \pi^{\pm}$ combinations versus the $\etapipi$ mass,
c)~the $\eta \pi^{\pm}$ mass projection of the $\fa$ region and
d)~of its sidebands.
}
\label{fig:a0f1}
\end{figure}

To search for the decay $\fa\ra\a0(980)\pi$ we select only data with $\spt > 0.1\GeV^2$. 
The corresponding \etapipi{} mass spectrum is shown in Figure~\ref{fig:a0f1}a. 
In Figure~\ref{fig:a0f1}b, both $\eta\pi^{\pm}$ mass combinations are plotted versus the \etapipi{} mass.
An accumulation of events with $\eta\pi^{\pm}$ mass around 0.98~\GeV{}
is observed correlated with the \fa{}.
The $\a0(980)$ signal is evident in Figure~\ref{fig:a0f1}c, where the \fa{} mass region is selected, 
$1.22<M(\etapipi)<1.34\GeV$.
No signal is observed in the sideband regions $1.12<M(\etapipi)<1.22\GeV$ and 
$1.34<M(\etapipi)<1.41\GeV$, Figure~\ref{fig:a0f1}d.
In order to evaluate the $\a0\pi$ contribution to the \fa{} signal, 
the $\eta\pi^{\pm}$ spectrum is fitted with
a resonance plus a background function as shown in Figure~\ref{fig:a0f1}c. 
The resonance is the convolution of a Breit-Wigner with a Gaussian resolution with width 0.014~\GeV{},
estimated from Monte Carlo.
The background function is obtained from the \fa{} sidebands of Figure~\ref{fig:a0f1}d.
The fit gives $M=0.985\pm 0.004\stat\pm 0.006\sys\GeV$, 
$\Gamma = 0.050\pm 0.013\stat\pm 0.004\sys\GeV$ and $318\pm 47\stat\pm 29\sys$ events. 
The fitted mass is in good agreement with the world average $M=0.9852\pm 0.0015\GeV$~\cite{PDG}.
The systematic uncertainties are obtained from the variation of the \fa{} and sideband mass limits
and from variation of the \spt{} cut.
A fit to the corresponding \etapipi{} mass spectrum of Figure~\ref{fig:a0f1}a 
gives $313\pm 29\stat\pm 6\sys$ events in the \fa{} peak,
where the systematic uncertainty is due to background subtraction.
Thus the observed number of \fa{} events is compatible with 100\% decay into $\a0\pi$.
Taking into account the statistical and systematic uncertainties, 
the measured branching fraction $\Gamma\bigl(\fa\ra\a0\pi\bigr) / \Gamma\bigl(\fa\ra\eta\pi\pi\bigr)$ 
is found to be greater than 0.69 at 95\% confidence level.

%
\newpage
\section*{Author List}
\typeout{   }     
\typeout{Using author list for paper 244 -- 246 }
\typeout{$Modified: Jul 31 2001 by smele $}
\typeout{!!!!  This should only be used with document option a4p!!!!}
\typeout{   }
%
%
%
%
%
%

\newcount\tutecount  \tutecount=0
\def\tutenum#1{\global\advance\tutecount by 1 \xdef#1{\the\tutecount}}
\def\tute#1{$^{#1}$}
\tutenum\aachen            
\tutenum\nikhef            
\tutenum\mich              
\tutenum\lapp              
\tutenum\basel             
\tutenum\lsu               
\tutenum\beijing           
\tutenum\berlin            
\tutenum\bologna           
\tutenum\tata              
\tutenum\ne                
\tutenum\bucharest         
\tutenum\budapest          
\tutenum\mit               
\tutenum\panjab            
\tutenum\debrecen          
\tutenum\florence          
\tutenum\cern              
\tutenum\wl                
\tutenum\geneva            
\tutenum\hefei             
\tutenum\lausanne          
\tutenum\lyon              
\tutenum\madrid            
\tutenum\florida           
\tutenum\milan             
\tutenum\moscow            
\tutenum\naples            
\tutenum\cyprus            
\tutenum\nymegen           
\tutenum\caltech           
\tutenum\perugia           
\tutenum\peters            
\tutenum\cmu               
\tutenum\potenza           
\tutenum\prince            
\tutenum\riverside         
\tutenum\rome              
\tutenum\salerno           
\tutenum\ucsd              
\tutenum\sofia             
\tutenum\korea             
\tutenum\utrecht           
\tutenum\purdue            
\tutenum\psinst            
\tutenum\zeuthen           
\tutenum\eth               
\tutenum\hamburg           
\tutenum\taiwan            
\tutenum\tsinghua          

{
\parskip=0pt
\noindent
{\bf The L3 Collaboration:}
\ifx\selectfont\undefined
 \baselineskip=10.8pt
 \baselineskip\baselinestretch\baselineskip
 \normalbaselineskip\baselineskip
 \ixpt
\else
 \fontsize{9}{10.8pt}\selectfont
\fi
\medskip
\tolerance=10000
\hbadness=5000
\raggedright
\hsize=162truemm\hoffset=0mm
\def\r{\rlap,}
\noindent

P.Achard\r\tute\geneva\ 
O.Adriani\r\tute{\florence}\ 
M.Aguilar-Benitez\r\tute\madrid\ 
J.Alcaraz\r\tute{\madrid,\cern}\ 
G.Alemanni\r\tute\lausanne\
J.Allaby\r\tute\cern\
A.Aloisio\r\tute\naples\ 
M.G.Alviggi\r\tute\naples\
H.Anderhub\r\tute\eth\ 
V.P.Andreev\r\tute{\lsu,\peters}\
% ...
\newpage
\begin{list}{A}{\itemsep=0pt plus 0pt minus 0pt\parsep=0pt plus 0pt minus 0pt
                \topsep=0pt plus 0pt minus 0pt}
\item[\aachen]
 I. Physikalisches Institut, RWTH, D-52056 Aachen, FRG$^{\S}$\\
 III. Physikalisches Institut, RWTH, D-52056 Aachen, FRG$^{\S}$
\item[\nikhef] National Institute for High Energy Physics, NIKHEF, 
     and University of Amsterdam, NL-1009 DB Amsterdam, The Netherlands
\item[\mich] University of Michigan, Ann Arbor, MI 48109, USA
\item[\lapp] Laboratoire d'Annecy-le-Vieux de Physique des Particules, 
     LAPP,IN2P3-CNRS, BP 110, F-74941 Annecy-le-Vieux CEDEX, France
\item[\basel] Institute of Physics, University of Basel, CH-4056 Basel,
     Switzerland
\item[\lsu] Louisiana State University, Baton Rouge, LA 70803, USA
\item[\beijing] Institute of High Energy Physics, IHEP, 
  100039 Beijing, China$^{\triangle}$ 
\item[\berlin] Humboldt University, D-10099 Berlin, FRG$^{\S}$
\item[\bologna] University of Bologna and INFN-Sezione di Bologna, 
     I-40126 Bologna, Italy
\item[\tata] Tata Institute of Fundamental Research, Mumbai (Bombay) 400 005, India
\item[\ne] Northeastern University, Boston, MA 02115, USA
\item[\bucharest] Institute of Atomic Physics and University of Bucharest,
     R-76900 Bucharest, Romania
\item[\budapest] Central Research Institute for Physics of the 
     Hungarian Academy of Sciences, H-1525 Budapest 114, Hungary$^{\ddag}$
\item[\mit] Massachusetts Institute of Technology, Cambridge, MA 02139, USA
\item[\panjab] Panjab University, Chandigarh 160 014, India.
\item[\debrecen] KLTE-ATOMKI, H-4010 Debrecen, Hungary$^\P$
\item[\florence] INFN Sezione di Firenze and University of Florence, 
     I-50125 Florence, Italy
\item[\cern] European Laboratory for Particle Physics, CERN, 
     CH-1211 Geneva 23, Switzerland
\item[\wl] World Laboratory, FBLJA  Project, CH-1211 Geneva 23, Switzerland
\item[\geneva] University of Geneva, CH-1211 Geneva 4, Switzerland
\item[\hefei] Chinese University of Science and Technology, USTC,
      Hefei, Anhui 230 029, China$^{\triangle}$
\item[\lausanne] University of Lausanne, CH-1015 Lausanne, Switzerland
\item[\lyon] Institut de Physique Nucl\'eaire de Lyon, 
     IN2P3-CNRS,Universit\'e Claude Bernard, 
     F-69622 Villeurbanne, France
\item[\madrid] Centro de Investigaciones Energ{\'e}ticas, 
     Medioambientales y Tecnol\'ogicas, CIEMAT, E-28040 Madrid,
     Spain${\flat}$ 
\item[\florida] Florida Institute of Technology, Melbourne, FL 32901, USA
\item[\milan] INFN-Sezione di Milano, I-20133 Milan, Italy
\item[\moscow] Institute of Theoretical and Experimental Physics, ITEP, 
     Moscow, Russia
\item[\naples] INFN-Sezione di Napoli and University of Naples, 
     I-80125 Naples, Italy
\item[\cyprus] Department of Physics, University of Cyprus,
     Nicosia, Cyprus
\item[\nymegen] University of Nijmegen and NIKHEF, 
     NL-6525 ED Nijmegen, The Netherlands
\item[\caltech] California Institute of Technology, Pasadena, CA 91125, USA
\item[\perugia] INFN-Sezione di Perugia and Universit\`a Degli 
     Studi di Perugia, I-06100 Perugia, Italy   
\item[\peters] Nuclear Physics Institute, St. Petersburg, Russia
\item[\cmu] Carnegie Mellon University, Pittsburgh, PA 15213, USA
\item[\potenza] INFN-Sezione di Napoli and University of Potenza, 
     I-85100 Potenza, Italy
\item[\prince] Princeton University, Princeton, NJ 08544, USA
\item[\riverside] University of Californa, Riverside, CA 92521, USA
\item[\rome] INFN-Sezione di Roma and University of Rome, ``La Sapienza",
     I-00185 Rome, Italy
\item[\salerno] University and INFN, Salerno, I-84100 Salerno, Italy
\item[\ucsd] University of California, San Diego, CA 92093, USA
\item[\sofia] Bulgarian Academy of Sciences, Central Lab.~of 
     Mechatronics and Instrumentation, BU-1113 Sofia, Bulgaria
\item[\korea]  The Center for High Energy Physics, 
     Kyungpook National University, 702-701 Taegu, Republic of Korea
\item[\utrecht] Utrecht University and NIKHEF, NL-3584 CB Utrecht, 
     The Netherlands
\item[\purdue] Purdue University, West Lafayette, IN 47907, USA
\item[\psinst] Paul Scherrer Institut, PSI, CH-5232 Villigen, Switzerland
\item[\zeuthen] DESY, D-15738 Zeuthen, 
     FRG
\item[\eth] Eidgen\"ossische Technische Hochschule, ETH Z\"urich,
     CH-8093 Z\"urich, Switzerland
\item[\hamburg] University of Hamburg, D-22761 Hamburg, FRG
\item[\taiwan] National Central University, Chung-Li, Taiwan, China
\item[\tsinghua] Department of Physics, National Tsing Hua University,
      Taiwan, China
\item[\S]  Supported by the German Bundesministerium 
        f\"ur Bildung, Wissenschaft, Forschung und Technologie
\item[\ddag] Supported by the Hungarian OTKA fund under contract
numbers T019181, F023259 and T024011.
\item[\P] Also supported by the Hungarian OTKA fund under contract
  number T026178.
\item[$\flat$] Supported also by the Comisi\'on Interministerial de Ciencia y 
        Tecnolog{\'\i}a.
\item[$\sharp$] Also supported by CONICET and Universidad Nacional de La Plata,
        CC 67, 1900 La Plata, Argentina.
\item[$\triangle$] Supported by the National Natural Science
  Foundation of China.
\end{list}
}
\vfill


\newpage
%

\vfill

\pagebreak

\begin{table}[ht]
\begin{center}
\begin{tabular}{|c|c|c|c|c|}
\hline
 & & & & \\[-2.5ex] 
\spt{} (\GeV{}$^2$)      & Events              & $M$ (\GeV{})     & $\epsilon$ (\%)         & $\Delta\sigma$ (pb) \\[0.2ex] \hline
$0.02-0.1$ \            & $ \ \,79\pm 22$  & $1.277\pm 0.007$ & $3.97\pm0.17\pm0.15$    & $28.9\pm\> 8.0\pm 2.7$ \\ 
$0.1\phantom{0}-0.4$  \            & $166\pm 22$         & $1.283\pm 0.004$ & $3.13\pm0.20\pm0.16$    & $57.7\pm\> 7.6\pm 5.3$ \\ 
$0.4\phantom{0}-0.9$  \            & $ \ \,91\pm 15$  & $1.287\pm 0.004$ & $3.35\pm0.29\pm0.25$    & $29.8\pm\> 4.7\pm 3.8$ \\ 
$0.9\phantom{0}-6.0$  \            & $ \ \,84\pm 11$  & $1.272\pm 0.004$ & $3.40\pm0.41\pm0.38$    & $28.2\pm\> 3.8\pm 4.3$ \\ \hline
\end{tabular}
\end{center}
\caption[]{Results of fits performed on the mass spectra of Figure~\ref{fig:f1bins}.
For each $\spt$ range the number of events in the peak, the mass $M$, 
the overall efficiency $\epsilon$  and the partial cross section $\Delta\sigma$ are presented. 
The uncertainties on the number of events and on the mass are statistical.
The uncertainties on the efficiency are respectively due to Monte Carlo statistics and
trigger behaviour. Statistical and systematic uncertainties on $\Delta\sigma$ are also presented.}
\label{tab:ptbins} 
\end{table}

\vspace{5mm}
\begin{table}[ht]
\begin{center}
\begin{tabular}{|c|c|c|c|c|}
\hline
 & & & & \\[-2.5ex] 
\spt{} (\GeV{}$^2$)   & Efficiency     & Background  & $\eta$ selection & Photon selection    \\[0.2ex] \hline
$0.02-0.1$  \           & \ 7.7        & 3.9         & 3.4              & 0.4                       \\
$0.1\phantom{0}-0.4$   \           & \ 8.1        & 3.4         & 2.6              & 0.5                       \\
$0.4\phantom{0}-0.9$   \           & 11.5         & 4.9         & 1.2              & 1.1                     \\
$0.9\phantom{0}-6.0$   \           & 14.8         & 2.7         & 1.2              & 0.2                    \\ \hline
\end{tabular}
\end{center}
\caption[]{Breakdown of the $\Delta\sigma$ systematic uncertainties, in \%, for each $\spt$ range,
as described in the text.}
\label{tab:syst} 
\end{table}

\pagebreak

\pagebreak

\pagebreak

\pagebreak

\pagebreak

\end{document}